\documentclass[useAMS,usenatbib]{mn2e}

\usepackage{graphicx}
\usepackage{rotating}

\newif\ifAMStwofonts                        
\newcommand{\lsimeq}{{_<\atop^{\sim}}}
\newcommand{\gsimeq}{{_>\atop^{\sim}}}

\title[Dusty SFR Evolution and Semi-Analytic Models]{Star Formation in {\textit{Herschel}}\thanks{{\it Herschel} is an ESA space observatory with science instruments provided by European-led Principal Investigator consortia and with important participation from NASA}'s Monsters versus Semi-Analytic Models}
\author[C. Gruppioni, F. Calura, F. Pozzi et al.]
{\parbox{\textwidth}{\raggedright C. Gruppioni$^{1}$\thanks{E-mail: carlotta.gruppioni@oabo.inaf.it}, 
F. Calura$^{1}$, 
F. Pozzi$^{2}$, 
I. Delvecchio$^{2}$,
S. Berta$^{3}$,
G. De Lucia$^{4}$,
F. Fontanot$^{4}$,
A. Franceschini$^{5}$, 
L. Marchetti$^{6}$, 
N. Menci$^{7}$,
P. Monaco$^{8}$,
M. Vaccari$^{9}$
}\vspace{0.4cm}\\
$^{1}$Istituto Nazionale di Astrofisica - Osservatorio Astronomico di Bologna, via Ranzani 1, I--40127 Bologna, Italy.\\
$^{2}$Dipartimento di Fisica e Astronomia, Universit\`a di Bologna, viale Berti Pichat 6, I--40127 Bologna, Italy.\\
$^{3}$Max-Planck-Institut f\"{u}r Extraterrestrische Physik (MPE), Postfach 1312, D--85741 Garching, Germany.\\
$^{4}$Istituto Nazionale di Astrofisica - Osservatorio Astronomico di Trieste, Via Tiepolo 11, I--34143 Trieste, Italy. \\
$^{5}$Dipartimento di Astronomia, Universit\`a di Padova, vicolo dell'Osservatorio 3, I--35122 Padova, Italy.\\
$^{6}$Department of Physical Sciences, The Open University, Milton Keynes MK7 6AA, UK.\\
$^{7}$INAF Ð Osservatorio Astronomico di Roma, via di Frascati 33, I--00040 Monte Porzio Catone, Italy.\\
$^{8}$Dipartimento di Fisica, Sezione di Astronomia, via Tiepolo 11, I--34143 Trieste, Italy.\\
$^{9}$Astrophysics Group, Department of Physics, University of Western Cape, Bellville 7535, Cape Town, South Africa.
}
\begin{document}

\date{Accepted 2015, May 27. Received 2015, May 6; in original form 2015, March 18.}

\pagerange{\pageref{firstpage}--\pageref{lastpage}} \pubyear{2015}

\maketitle

\label{firstpage}

\begin{abstract}
We present a direct comparison between the observed star formation rate functions (SFRF) and the state-of-the-art predictions of semi-analytic models (SAM) of galaxy formation and evolution. 
We use the PACS Evolutionary Probe Survey (PEP) and Herschel Multi-tiered Extragalactic Survey (HerMES) data-sets in the COSMOS and GOODS-South fields, combined
with broad-band photometry from UV to sub-mm, to obtain total (IR$+$UV) instantaneous star formation rates (SFRs) for individual {\em Herschel} galaxies up to $z$$\sim$4, subtracted of
possible active galactic nucleus (AGN) contamination.
The comparison with model predictions shows that SAMs broadly reproduce the observed SFRFs up to $z$$\sim$2, when the observational errors on the SFR are taken into account. 
However, all the models seem to under-predict the bright-end
of the SFRF at $z$$\gsimeq$2. The cause of this underprediction could lie in an improper modelling of several model ingredients, like too strong (AGN or stellar) feedback in the brighter objects 
or too low fall-back of gas, caused by weak feedback and outflows at earlier epochs.
\end{abstract}

\begin{keywords}
cosmology: observations --  galaxies: evolution -- galaxies: formation -- galaxies: luminosity function -- galaxies: star-formation -- infrared: galaxies.
\end{keywords}

\section{Introduction}
The study of how the star formation rate (SFR) in galaxies evolves with redshift provides important constraints to the galaxy formation and evolution theories. In particular, semi-analytic models 
(SAMs; e.g. White \& Frenk 1991; Kauffmann et al. 1993; Springel et al. 2001; Monaco et al. 2007; Guo et al. 2011; Benson 2012; Menci et al. 2012; Henriques et al. 2013), need to be directly compared with observations to obtain insight of the relevant physical processes. 
The first and most popular SAMs are three, commonly named ``Munich'' (starting with the models of Kauffmann, White \& Guiderdoni 1993), ``Durham'' (beginning with the models of Cole et al. 1994), and ``Santa Cruz'' (beginning with the models of Somerville \& Primack 1999); more recent SAMs include, i.e., Croton et al. 2006; Bower et al. 2006; Somerville et al. 2008; Fontanot et al. 2009; Guo et al. 2011; Somerville et al. 2012. The main differences between these models lie in the prescriptions adopted for some of the most basic baryonic processes, such as star formation, gas cooling and feedback.
One of the processes that must be modelled and compared to data is the evolution of the SFR over the cosmic time.
However, the derivation of an accurate SFR from observational data is difficult, due to the many uncertainties involved in its reconstruction. An important source of uncertainty comes from dust extinction. 
The rest-frame ultraviolet (UV) light emitted by young and massive stars, strictly connected to the instantaneous SFR in galaxies, is strongly absorbed by dust, and re-radiated in the infrared (IR) bands. 
Dust attenuation, as well as other galaxy physical properties, evolves with cosmic time and shows a peak between $z$$\sim$1 and 2 (e.g. 
Burgarella et al. 2013). Knowing how dust attenuation evolves in redshift is therefore crucial to study the redshift evolution of the SFR: to this purpose, combining UV information with 
direct observations in the IR region is probably the best tool to account for the total SFR (e.g., \citealt{kenn98}). In fact, IR surveys covering a wide range 
of redshifts are extremely useful to estimate the global IR luminosity, since they provide a direct measurement of the amount of energy absorbed and re-emitted by dust (e.g., what is missed by UV surveys). 

{\em Herschel} (\citealt{pilb10}), with its 3.5-m mirror, has been the first telescope which allowed us to detect the far-IR population to high redshifts ($z$$\sim$3--4) and to derive its rate of evolution through a detailed LF analysis (\citealt{grup13}; \citealt{mag13}) thanks to the extragalactic surveys provided by the {\em Photodetector Array Camera \& Spectrometer} ({\em PACS}; \citealt{pog10}) and {\em Spectral and Photometric Imaging Receiver} ({\em SPIRE}; \citealt{grif10}) in the far-IR/sub-mm domain (i.e. PACS Evolutionary Probe, PEP, \citealt{lutz11}; {\it Herschel} Multi-tiered Extragalactic Survey, HerMES, \citealt{oliv12};
GOODS-{\it Herschel}, \citealt{elb11}; {\it Herschel}-ATLAS, \citealt{eal10}). 
PEP and HerMES are the major {\em Herschel} Guaranteed Time extragalactic key-projects, designed specifically to determine the cosmic evolution of dusty star formation and of the IR LF, 
and include the most popular and widely studied extragalactic fields with extensive multi-wavelength coverage available (deep optical, near-IR and {\em Spitzer} imaging and spectroscopic and photometric redshifts ; see \citealt{berta11}; \citealt{lutz11}; \citealt{oliv12} for a detailed description of the fields and observations). 
The far-IR domain in galaxies, although potentially contaminated by the presence of an AGN, has been probed to be dominated by star formation (i.e. \citealt{hatz10}; \citealt{delv14}). 
Therefore, PEP and HerMES, and all the ancillary data available in the fields, give us the opportunity to disentangle star formation from AGN contribution and to study in detail the 
evolution of the SFR with cosmic time since the Universe was about a billion years old. \\
In a recent paper from Pozzi et al. (2015),  the observed IR PEP/HerMES LFs
have been reproduced by means of a phenomenological model considering two galaxy populations characterised by different evolutions,
 i.e. a populations of late-type sources and a populations of proto-spheroids. In the model, the
IR luminosity functions (linked to the SFR) have been reproduced, as well as the literature
K-band luminosity functions (directy linked to the stellar mass), showing that most of the PEP-selected sources
observed at $z$$>$2 can be explained as progenitors of local spheroids caught during their formation.\\
Similar evolutionary rates have been found by \citet{grup13}, deriving the far- and total IR (i.e. rest-frame 8--1000 $\mu$m) LFs 
from the {\em Herschel} data obtained within the PEP and HerMES projects up to $z$$\sim$4. Since a large fraction of {\em Herschel} selected objects have been found to contain an AGN
(\citealt{berta13}; \citealt{grup13}; \citealt{delv14}), an accurate quantification of the AGN contribution to the IR luminosity is needed in order to derive reliable SFRs (e.g., not contaminated by
AGN activity) from these sources. 
\citet{delv14}, through a detailed SED decomposition analysis (see \citealt{berta13}), have disentangled the contribution to the total IR luminosity due to AGN activity and that due to SF for the 
whole PEP sample. By starting from the work of \citet{delv14}, but considering the contribution of SF only, in the present paper we focus on the determination of the SFR function (SFRF) and 
SFR density (SFRD) and compare the results with the predictions of state-of-the-art SAMs of galaxy formation and evolution.

This paper is organised as follows. We discuss the PEP multi-wavelength catalogue in Section 2 and the theoretical predictions and comparison with data in Section 3; finally we present our conclusions in Section 4. 

\noindent Throughout this paper, we use a Chabrier (2003) initial mass function (IMF) and we assume a $\Lambda$CDM cosmology with $H_{\rm 0}$\,=\,71~km~s$^{-1}$\,Mpc$^{-1}$, $\Omega_{\rm m}$\,=\,0.27, and $\Omega_{\rm \Lambda}\,=\,0.73$ for data derivations. Note that, although not affecting the results of this paper (see, e.g., \citealt{wang08}), the considered SAMs use slightly different cosmologies.

\section[]{The SFR function of \textit{Herschel} selected galaxies}
\subsection{The data-set}
We have considered the \textit{Herschel} PACS (70, 100 and 160 $\mu$m) and SPIRE (250, 350 and 500 $\mu$m) data in the COSMOS and GOODS-S fields from the PEP and HerMES Surveys and all the multi-wavelength
data-set associated to the far-IR sources. The reference sample is the PEP blind catalogue selected at 160 $\mu$m to the 3$\sigma$ level, which consists of 4118 and 492 sources 
respectively in COSMOS (to 10.2 mJy in $\sim$2 deg$^2$) and GOODS-S (to 1.2 mJy in $\sim$196 arcmin$^2$. 
As described in detail by \citet{berta11} and \citet{grup13}, our sources have been associated to the ancillary catalogues by means of a multi-band likelihood ratio technique (e.g., \citealt{suth92}), starting from the longest available wavelength (160\,$\mu$m, PACS) and progressively matching 100\,$\mu$m (PACS), 70\,$\mu$m (PACS, GOODS-S only) and 24\,$\mu$m ({\em Spitzer}/MIPS). 
In the GOODS-S field, we have associated to our PEP sources the 24-$\mu$m catalogue by \citet{mag09} (extracted with IRAC 3.6-$\mu$m positions as priors), that we have matched with the optical$+$near-IR$+$IRAC MUSIC catalogue of \citet{graz06}, revised by \citet{sant09}, which includes spectroscopic and photometric redshifts. In COSMOS, we have matched our catalogue with the deep 24-$\mu$m sample of \citet{lef09} and with the IRAC-based catalogue of \citet{ilb10}, including optical and near-IR photometry and photometric redshifts. In HerMES a prior source extraction was performed using the method presented in \citet{rose10}, based on MIPS-24\,$\mu$m positions. The 24-$\mu$m sources used as priors for SPIRE source extraction are the same as those associated with our PEP sources through the likelihood ratio technique.
We have therefore associated the HerMES sources with the PEP sources by means of the 24-$\mu$m sources matched to both samples. For most of our PEP sources ($\sim$87 per cent) we found a $>$3$\sigma$ SPIRE counterpart in the HerMES catalogues.
Redshifts (either spectroscopic or photometric) are available for all the sources in GOODS-S and for 93\% of the COSMOS sample
(references and details also in \citealt{berta11} and \citealt{grup13}).
\begin{figure*}
\includegraphics[width=17cm,height=10cm]{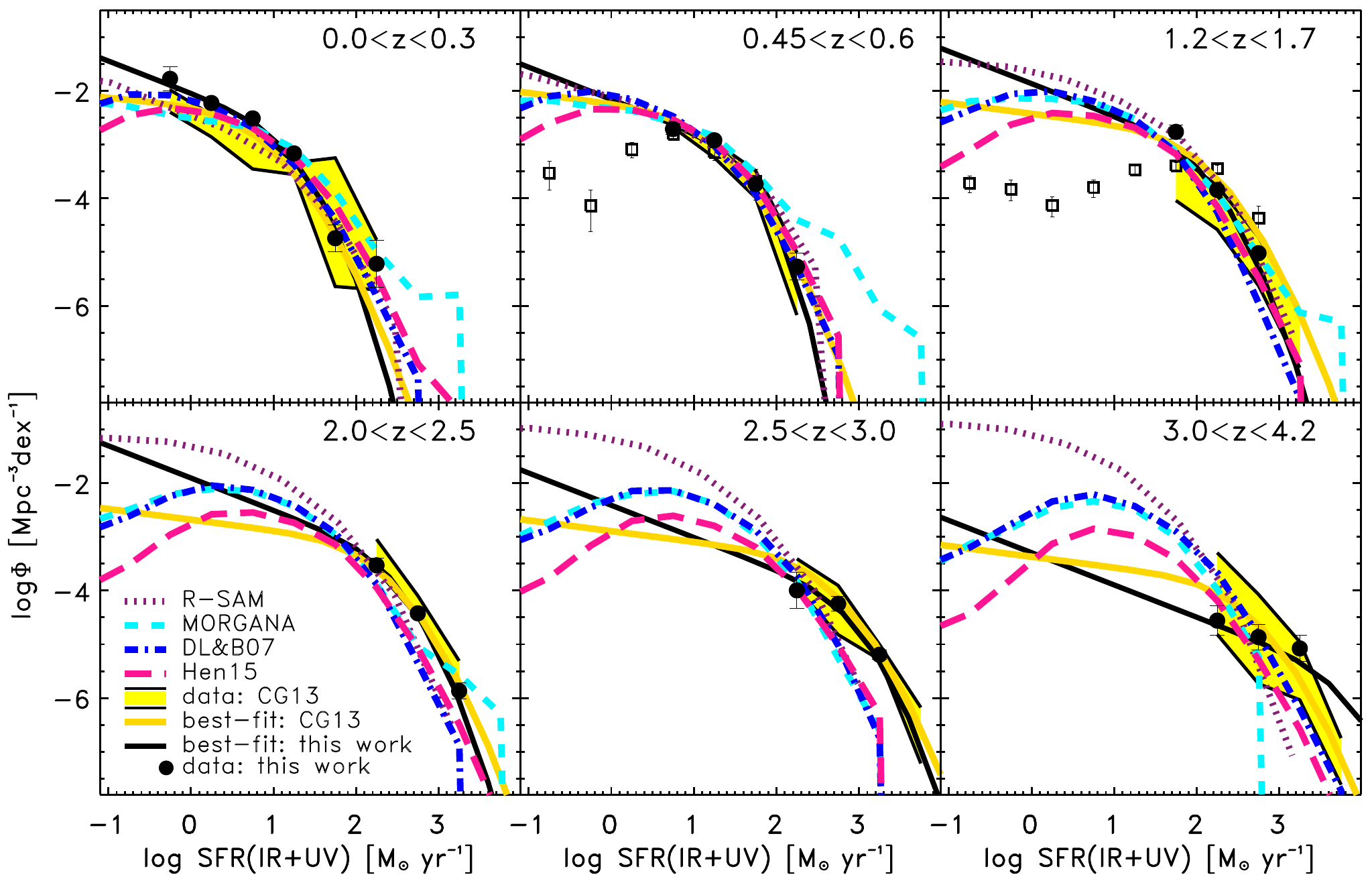}
\caption{IR$+$UV SFRF estimated through the 1/V$_{\rm max}$ method in 6 representative redshift bins, by combining the data from the PEP GOODS-S and COSMOS fields using the \citet{avni80} technique (black filled circles). The black solid line represents our best fit to our data with a modified Schechter function, while the yellow solid line is the total 
IR LF (without excluding AGN contribution through SED decomposition) obtained by \citet{grup13}, converted to a SFRF. The SFRFs of 24-$\mu$m sources with log($M$/$M_{\odot}$)$>$10 in the GOODS-S by  \citet{font12} are plotted for comparison as black open squares.
The SAMS predictions are shown as purple dotted (R-SAM), sea-green short-dashed (MORGANA), blue dot-dashed (\citealt{delucia07}) and deep-pink long-dashed (\citealt{hen15})  coloured lines.}
\label{figSFRLFbol}
\end{figure*}
\subsection{The SFR Function and SFR Density}
\citet{grup13} derived the far- and total IR (i.e. rest-frame 8--1000 $\mu$m) LFs from the {\em Herschel} data obtained within the PEP and HerMES projects up to $z$$\sim$4.  
To compute the SFRF, we have used the same data and method used by \citet{grup13} for deriving the total IR LF, but here we have subtracted - for each source individually -
the AGN contribution from each SED, as estimated by \citet{delv14}, to obtain the IR luminosity due to SF only.
In order to disentangle the possible AGN contribution from that related to the host galaxy, \citet{delv14} have performed a broad-band SED decomposition of our PEP sources 
using the MAGPHYS code (\citealt{dacun08}), which is a public code using physically motivated 
templates to reproduce the observed galaxy SEDs, as modified by \citet{berta13} to include also the AGN component (from \citealt{fritz06} and \citealt{feltre12} models).
\citet{delv14} found significant (at 99 per cent) contribution from AGN in 37 per cent of the PEP sources and used the 
IR luminosity of the AGN component to derive the AGN bolometric LF and the SMBH growth rate across cosmic time up to $z$$\sim$3. Note that  \citet{delv14} choose to derive the
SMBH accretion function only up to $z$$\sim$3, based on the fact that too few BH accretion data points were available at $z$$>$3 in order to provide an acceptable fit. 
On the contrary, in this work, we subtract the AGN component from the total IR luminosity of the sources with a significant AGN activity to obtain the contribution due
to SF only ({\em L}$_{\rm IR}^{\rm SF}$). In contrast with the work of \citet{delv14}, 
we can estimate the SFRF also in the 3$<$$z$$<$4.2 interval, since the SFR data above the completeness limit allowed us to obtain an acceptable fit (though
with larger uncertainties than at lower $z$). 
We have used the same calibration of Santini et al. (2009) and Papovich et al. (2007) to estimate the total instantaneous SFR (then used to derive the SFRF):   
\begin{equation}
SFR_{\rm IR+UV}/{\rm M}_{\odot} {\rm yr}^{-1}=1.8\times10^{-10}\times(2.2\times L_{\rm UV}+L_{\rm IR})
\label{eq:SFR}
\end{equation}
\noindent with $L_{\rm UV}$$=$1.5$\times$$L_{\rm 2700 \AA}$ computed from the best-fit template SED, and $L_{\rm IR}=${\em L}$_{\rm IR}^{\rm SF}$. 
To derive the SFRF, we have used the 1/V$_{\rm max}$ method (\citealt{schm68}), combining the data in the two fields following \citet{avni80}. 
According to this method, the SFRF value and its uncertainty in each SFR bin have been computed as:
\begin{equation}
\centering
\Phi(SFR,z)=\frac{1}{\Delta SFR}\big[\sum_{\rm i}{\frac{1}{w_{\rm i} \times V_{\rm max,i}}}\pm \sqrt {\sum_{\rm i}{\frac{1}{(w_{\rm i} \times V_{\rm max,i})^2}}}\big]
\end{equation}
where $V_{\rm max,i}$ is the comoving volume over which the $i$-${\rm th}$ galaxy could be observed, $\Delta SFR$ is the size of the SFR bin (in logarithmic scale), and $w_{\rm i}$ is the completeness correction factor of 
the $i$-${\rm th}$ galaxy. These completeness correction factors are a combination of the completeness corrections given by \citet{berta11}, 
derived as described in \citet{lutz11}, to be applied to each 
source as a function of its flux density, together with a correction for redshift incompleteness. Additional details are given in \citet{grup13}.

The resulting SFRFs in different redshift intervals are shown in Fig.~\ref{figSFRLFbol} (black solid circles) and presented in Table~\ref{TabSFRF} (together with their associated 1-$\sigma$ uncertainties). 
For comparison, the SFRF of sources with log($M$/M$_{\odot}$)$>$10 in the GOODS-S derived by \citet{font12} have been also plotted, although the two samples have different 
selections (i.e. the \citealt{font12}
sample is selected at 24 $\mu$m and complete in mass, while ours is flux-limited at 160-$\mu$m), therefore are not directly comparable at the fainter SFRs (affected by the sample cuts). 
Our best-fit solution with a modified Schechter function (\citealt{saund90}) has also been reported (black solid line)
and compared with the best-fit to the total IR LF of \citet{grup13} (yellow solid line), converted to SFRF through the \citet{kenn98} relation scaled
to the Chabrier IMF (although containing also the AGN contribution). 

Note that, apart from the two lower redshift 
bins ($z$$<$0.45), the faint-end of our SFRFs is not constrained by data, therefore we can only derive the value of $\alpha$ at low-$z$, then fix it
and keep that value also in the higher redshift bins. This Hobson's choice implies the assumption that the faint-end slope does not vary with redshift. 
However, we note that at $z$$<$0.45, where $\alpha$ is constrained by data, the SFRF obtained from the conversion of the total IR LF of \citet{grup13} is flatter at low SFRs than
the (IR$+$UV) SFRF (e.g. the yellow uncertainty area is lower than the faintest SFR data point). 
This can be interpreted as due either to a major contribution of the UV to the faint-end (almost negligible at higher SFRs, dominated by the IR)
or/and to the fact that some sources might move to fainter luminosity bins when the AGN contribution is removed,
thus steepening the SFRF. 
\begin{figure}
\includegraphics[width=8cm]{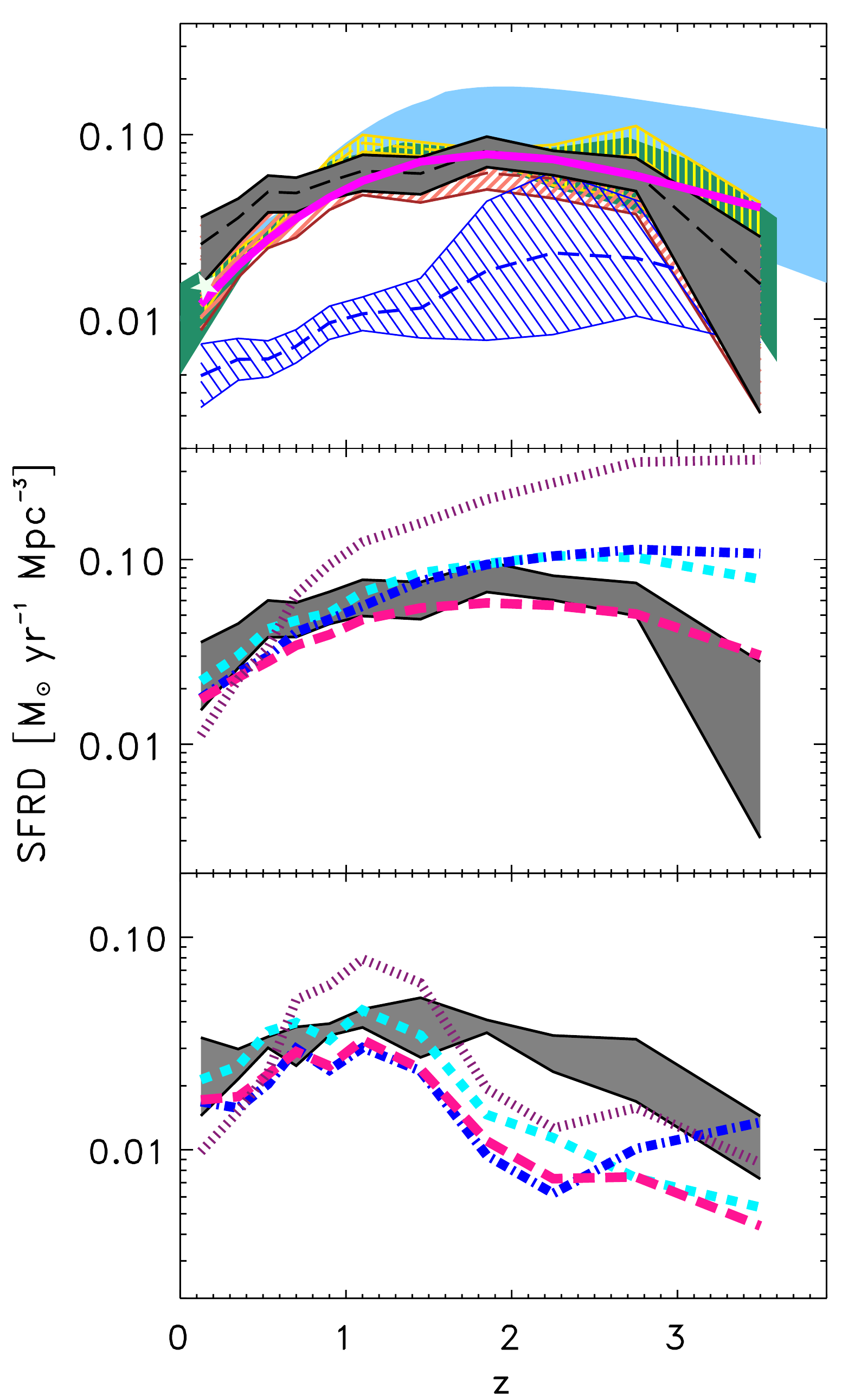}
\caption{Redshift evolution of the comoving SFRD. The results of integrating the best-fitting curve for our observed IR$+$UV SFRD in each $z$-bin is shown as grey filled area ($\pm$1$\sigma$ uncertainty locus). {\em Top panel:} Our IR$+$UV SFRD estimate is compared with other derivations: the yellow line-filled area shows the total IR Luminosity Density resulting from integrating the \citet{grup13} best-fitting curve (i.e. converted to SFRD without excluding the AGN contribution); the orange line-filled area shows the uncertainty region of the only-IR SFRD from our data (after AGN removal); the blue line-filled area is the uncorrected for extinction UV SFRD by \citet{cuc12}; the dark green filled area is the \citet{burg13} estimate from {\em Herschel} data; the pale blue filled area represents the fit to optical/UV data by \citet{behr13}; the magenta line is the the best-fitting function to IR and UV comoving SFRD by Madau \& Dickinson (2014). The honeydew star shows the SFRD value obtained by Marchetti et al. (2015) from the HerMES local wide area sample, by
combining the IR and the UV SFRs (but without excluding the AGN contribution). {\em Middle panel:} Our IR$+$UV SFRD estimate shown in the previous panel is compared with the SAMs predictions (same colours and line-styles as in the previous figures) integrated over the same range of SFRs (from $log_{10}(SFR)$$=$$-$1.5). {\em Bottom panel:} same as in the previous panel, but with our best-fitting function and SAMs
integrated only over the SFR range covered by our data.}
\label{figsfrz}
\end{figure}
The latter explanation is also consistent with our finding that in all the redshift bins but the highest one (which is also the most uncertain, due to the high fraction of photometric redshifts), 
the bright-end of the SFRF (black solid line) is always steeper than that of the total IR LF converted to SFRF (yellow solid line). Since the AGN-dominated sources contribute mainly to the bright-end 
of the IR LF (see \citealt{grup13}), 
this difference is due to the subtraction of the AGN component from their SEDs.

The SFRF of 24-$\mu$m selected sources with $M$$>$10$^{10}$ M$_{\odot}$ by \citet{font12} is in good agreement with our determination, in the common redshift and SFR range, 
with the \citet{font12} one being flatter in the lower SFR common bin, likely due to the mass cut in their sample.  
\begin{table*}
\begin{minipage}{180mm}

\caption{PEP SFRF and SFRD}\label{TabSFRF}

\scriptsize
\begin{tabular}{|c|cccccccc|cc|}
\hline  \hline
               & \multicolumn{8}{c}{log(${\Phi}$/Mpc$^{-3}$ dex$^{-1}$)} & $SFRD_{\rm IR+UV}$  & $SFRD_{\rm IR}$ \\
               & \multicolumn{8}{c}{}  &\multicolumn{2}{c}{(M$_{\odot}$ yr$^{-1}$ Mpc$^{-3}$)} \\ \hline
                z & \multicolumn{8}{c}{log($SFR_{\rm IR+UV}$/M$_{\odot}$ yr$^{-1}$)} &  \\ \hline \hline
                 & $-$0.5$\div$0.0      &    0.0$\div$0.5        &     0.5$\div$1.0        &      1.0$\div$1.5        &      1.5$\div$2.0         &    2.0$\div$2.5        &     2.5$\div$3.0       &3.0$\div$3.5  & \\ \hline
0.0$-$0.3 &  $-$1.78$\pm$0.23  &$-$2.26$\pm$0.05  &$-$2.53$\pm$0.04  &$-$3.29$\pm$0.05   &  $-$4.92$\pm$0.31  &$-$5.22$\pm$0.43  &                                 &                          &   0.025$\pm$0.005 &0.022$\pm$0.016\\ 
0.3$-$0.45 &                                 &$-$2.52$\pm$0.14 & $-$2.36$\pm$0.11 & $-$2.85$\pm$0.08 & $-$4.29$\pm$0.11  &  $-$5.22$\pm$0.31  &                                        &                         &   0.035$\pm$0.010   &0.028$\pm$0.012\\
0.45$-$0.6 &                                &                                  &$-$2.70$\pm$0.09 &$-$2.94$\pm$0.06 &$-$3.79$\pm$0.05 & $-$5.45$\pm$0.31   &                                        &                         &   0.049$\pm$0.014 & 0.038$\pm$0.013\\
0.6$-$0.8  &                                 &                                   &$-$2.38$\pm$0.19 &$-$2.67$\pm$0.06 &$-$3.51$\pm$0.03 &  $-$4.96$\pm$0.13    &                                      &                          &  0.056$\pm$0.013 &0.039$\pm$0.015\\
0.8$-$1.0 &                                   &                                  &                                 &$-$3.01$\pm$0.08 & $-$3.26$\pm$0.04 & $-$4.40$\pm$0.06   & $-$6.17$\pm$0.43      &                         & 0.064$\pm$0.016 &0.050$\pm$0.019\\
1.0$-$1.2 &                                 &                                   &                                &$-$2.95$\pm$0.11  & $-$3.17$\pm$0.08 &  $-$4.13$\pm$0.04     &                                       &                        &  0.062$\pm$0.014 &0.056$\pm$0.026\\ 
1.2$-$1.7 &                                &                                   &                                &                                     & $-$2.76$\pm$0.13 &  $-$3.90$\pm$0.04   & $-$5.23$\pm$0.08    &                          & 0.082$\pm$0.021 &0.051$\pm$0.025\\
1.7$-$2.0 &                                 &                                   &                               &                                     &                               & $-$3.96$\pm$0.10   &  $-$4.61$\pm$0.05   &  $-$6.56$\pm$0.43 &0.071$\pm$0.019 &0.062$\pm$0.023\\
2.0$-$2.5&                                  &                                   &                               &                                     &                             &$-$3.54$\pm$0.13 & $-$4.46$\pm$0.06  & $-$6.09$\pm$0.19 &0.062$\pm$0.021 &0.058$\pm$0.015\\  
2.5$-$3.0 &                                &                                   &                               &                                     &                             &$-$4.00$\pm$0.34 & $-$4.23$\pm$0.13 & $-$5.32$\pm$0.10  &  0.056$\pm$0.020 &0.053$\pm$0.016\\
3.0$-$4.2  &                                &                                   &                               &                                     &                             &$-$4.67$\pm$0.28 & $-$4.94$\pm$0.20 & $-$5.34$\pm$0.32 &  0.028$\pm$0.012 &0.016$\pm$0.012\\  
\hline \hline
\end{tabular}
\end{minipage}
\end{table*}

By integrating the best-fitting modified Schechter function to our IR$+$UV (IR) SFRFs down to $log_{10}(SFR)$$=$$-$1.5 in the different redshift bins, from $z$$\sim$0 to $z$$\sim$4, we have derived the comoving IR$+$UV (IR) SFR density 
(SFRD), as presented in the 10$^{th}$ (last) column of Table~\ref{TabSFRF} and shown in Fig~\ref{figsfrz} as a grey-filled (orange line-filled) area. 
For comparison, other derivations from different bands are shown (i.e. the integration of the best-fitting Schechter function to the total IR LF, containing AGN, converted to SFR as yellow line-filled area; 
the IR$+$UV SFRD by \citealt{burg13} as dark-green filled area; the fit to optical/UV data by \citealt{behr13} as pale blue filled-area; the UV SFRD - uncorrected for extinction - derivation by \citealt{cuc12} as
blue line-filled area). 

From the comparison between our SFRD and previous derivations, we notice that the IR$+$UV SFRD estimated in this work is higher at low redshift (i.e. $z$$<$0.5), while it agrees within the uncertainties with the
other estimates at higher $z$. The IR-only SFRD, given the larger uncertainties, is consistent with the optical SFRD by \citet{behr13} and with the \citet{grup13} IR LD at low-$z$, although the
average value is also higher than the latter estimates. Therefore, the low redshift difference is likely due to the UV SFR contribution (that in percentage is higher at low-$z$), but mostly to the AGN contribution
subtraction. The (uncorrected for dust extinction) UV SFRD by \citet{cuc12} is significantly lower than our IR or IR$+$UV one over the 0$<$$z$$<$3 range, while at $z$$>$3 it becomes comparable. 
This is consistent with the peak of dust extinction being around $z$$\sim$1.5-2, with dust attenuation rapidly decreasing at higher redshifts ($>$3-4; e.g., \citealt{burg13}). 
In  the {\em top panel} of Fig.~\ref{figsfrz} we also show the redshift evolution of the total SFRDs as obtained (from the same data sample) by Burgarella et al. (2013; dark-green shaded region). The differences (i.e. the \citealt{burg13} estimate is lower than the current one at $z$$<$1 and slightly higher, but still within the uncertainties, at $z$$>$2.5) can be ascribed to the fact that in the previous
analysis an average AGN contribution for each population had been subtracted from the total IR luminosity density (then converted to SFRD and summed to the UV SFRD), while in this work 
an accurate object-by-object subtraction of the IR luminosity contribution due to the AGN has been performed, thanks to the detailed SED-fitting and decomposition of \citet{delv14}, and a proper 
SFRF has then been calculated from the obtained IR$+$UV SFRs. 

Finally, we note that, while the best-fitting function to the comoving SFRD from IR and UV data by Madau \& Dickinson (2014) is in good agreement
with our derivation (although slightly lower at $z$$<$0.5 and higher at $z$$>$3), at $z$$>$0.5 the average \citet{behr13} estimate is always higher (although consistent within the large uncertainty region),
maybe due to large extinction corrections.

Finally, in Fig.~\ref{figsfrz} we show the SFRD value obtained by Marchetti et al. (2015) from the HerMES wide-area sample, by combining the SFR from IR and UV (in analogy with the present work),
but without excluding the AGN contribution. The value is in good agreement with the result of \citet{grup13} (and the others from the literature) at the same redshift and only marginally consistent with out IR$+$UV SFRD result.
\begin{figure}
\includegraphics[width=8cm,height=11cm]{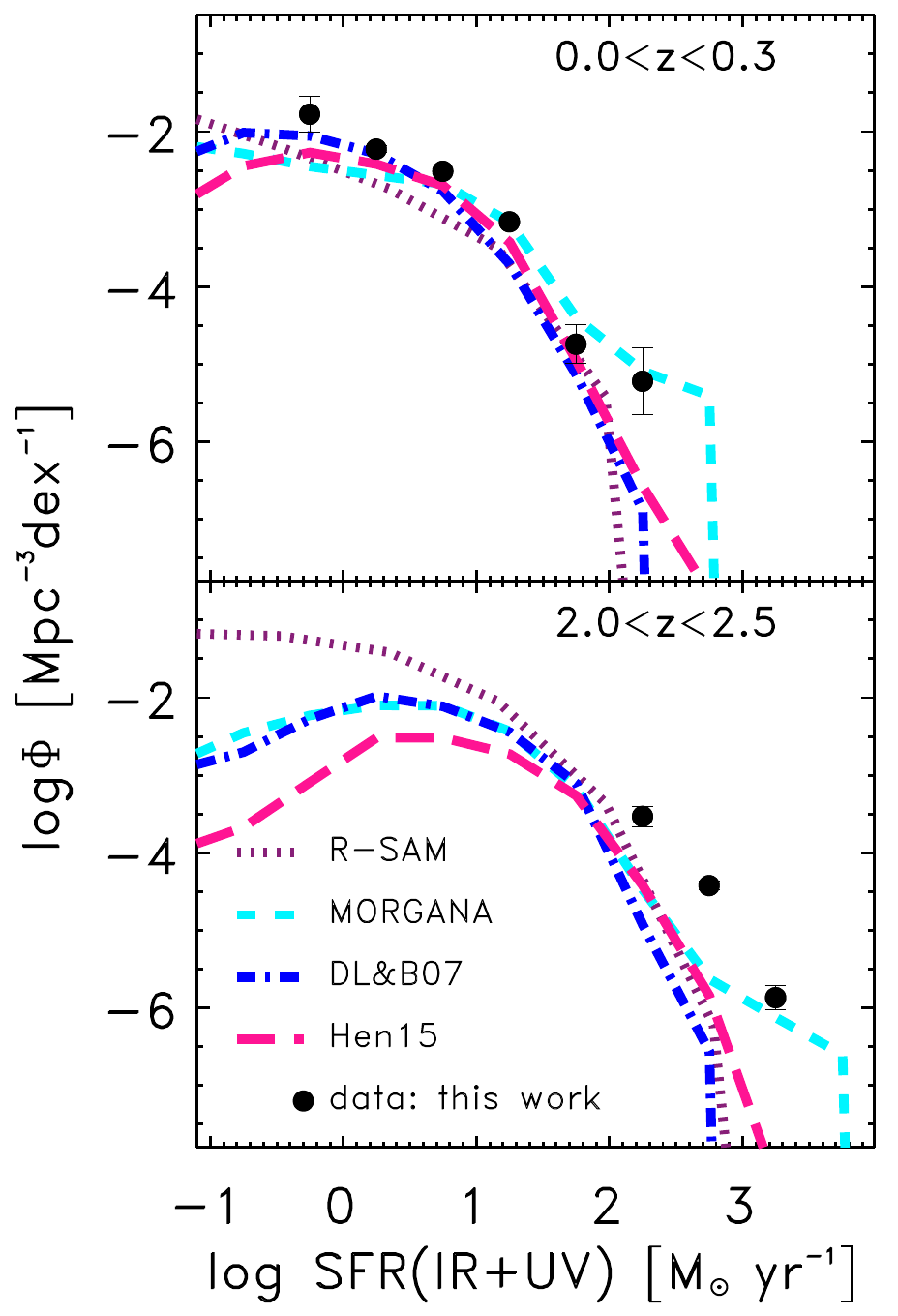}
\caption{Total (IR$+$UV) SFRF (same as in Fig.~\ref{figSFRLFbol}; black filled circles) in two representative redshift bins compared to SAMs predictions not convolved with an error of 0.3 dex (same colours and line-styles as in Fig.~\ref{figSFRLFbol}).}
\label{figSFRLFnoconv}
\end{figure}

\section[]{Semi-analytical model predictions}
In this section, we compare the observed SFRFs with results obtained
with four SAMs of galaxy formation. The SAMs considered in this work are:
the MOdel for the Rise of Galaxies aNd Agns (MORGANA, \citealt{mon07}; sea-green short dashed lines in the
figures), R-SAM (Menci et al. 2012, 2014; purple dotted lines) and 
two versions of the Munich model\footnote{obtained from the publicly available database\\ http://gavo.mpa-garching.mpg.de/Millennium/} by De Lucia \& Blaizot (2007; 
dot-dashed blue lines) and Henriques et al. (2015; deep-pink long dashed lines).
Note that these SAMs use slightly different cosmologies (see the relative papers for details), although this does not affect the 
results discussed in this paper (as shown by \citealt{wang08}). Moreover, all the SAMs' predictions
(except the R-SAM's ones) are mass-limited, with a cut at M$=$10$^9$ M${\odot}$, though this
selection affects only the low-SFRs, typically lower than those reached by our data.
Since these models have not been "tuned" to reproduce the SFRFs, the results shown in this work
can be considered as genuine SAMs predictions.

SAMs treat the physical processes involving baryons (thermal state and
infall/outflow of gas, star formation, feedback, accretion onto a
black hole, AGN feedback) within the backbone of dark matter (DM) halo
merging histories, produced by the gravitational collapse of DM (see
Benson 2001 for a review).  We refer the reader to the papers cited above for all details 
on the models.

Together with the observational SFRF, Fig.~1 reports, at the six
chosen redshift ranges, the SFRFs predicted by the four SAMs.  To take
into account the errors in the observational determination of SFRs, model
predictions have been convolved with a fiducial error of 0.3 dex (assuming a log-normal error distribution 
for the SFR with an amplitude of 0.3 dex).
As discussed by Fontanot et al. (2009; 2012), this value is roughly equal to the median formal 
error of SFRs in the GOODS-MUSIC catalogue (Santini et al. 2009) and allows us to 
determine the gross effect of (random) uncertainties in SFR determinations.
We consider this value suitable also for our far-IR-based SFR.

At low redshift ($z$$<$1.7, the three upper panels), all but MORGANA model predictions,
are remarkably similar on the bright end. The MORGANA model at $z$$\lsimeq$0.8 shows an
excess of very bright sources, that is connected to the
specific model of radio-mode AGN feedback, where accretion onto the
central black hole takes place from the cold, star forming gas in the bulge, so the
suppression of star formation is partial. 
All models tend to give a power-law tail for the 
bright-end of the SFRF, that is broadly compatible with the analytical extrapolation 
of the observed SFRF presented in Section 2. 
The broadly good agreement between models and data at the bright-end
breaks in the very important redshift range from 2 to 3, with the very steep SFRFs
produced by all models becoming apparently different from the analytic fit to data
in the highest redshift bin. However, the slope of this analytic fit is
determined by $z=0-0.3$ data, while at $z\ga2$, there is clear
observational evidence of a steep (or even steepening of the) UV luminosity function 
(e.g. Cucciati et al. 2012). The consistency at least of one of these models (e.g. MORGANA) 
with the LF of Lyman-break galaxies has been investigated by Lo Faro et al. (2009).
At $z$$\gsimeq$2, the models underpredict the bright-end of the SFRF by a factor of
$\sim$0.2$-$0.3 dex.  
This result depends sensitively on the adopted
modelling of observational error.  Figure 3 shows the comparison of
models and data for two redshift bins at $z$$\sim$0 and $z$$\sim$2, {\em
without} convolving with 0.3 dex error.  While the comparison at
$z$$=$0 remains acceptable, the disagreement at $z\sim2$ worsens
dramatically.  Only extreme assumptions on the error on SFRs would
allow to recover the $z$$\simeq$2 SFRF.  

We stress that in the highest redshift bin (3$<$$z$$<$4.2), where the knee of the 
observed SFRF is not clearly detectable and the shape seems to change significantly, the 
fraction of photometric redshifts and of the power-law SED AGN (with an uncertain photo-$z$ determination)
is higher than at lower $z$'s. For this reason, this $z$-bin is more affected 
by uncertainties than the other ones and the discrepancy with model predictions here must be 
taken just as indications (to be furtherly verified).  

The inability of SAMs to
reproduce the bright-end of the {\em Herschel} LF at $z$$\gsimeq$2 had
previously been found by \citet{niemi12}, though based on very preliminary 
{\em Herschel} LF results. It might be
connected to the tendency of models to underestimate the main sequence
of star-forming galaxies at $z$$\sim$2 (e.g., \citealt{dutt10}).  
Because SFRs are determined by the complex pattern of gas inflows and outflows, 
in and out DM halos, and because inflows are determined by outflows taking place 
at previous times, identifying the cause of this disagreement is not easy. One 
possibility could be excessive feedback, possibly from AGN either in the radio 
mode (cooling and thus SFR is over-suppressed in these objects) or in the quasar 
mode (quasar-triggered outflows limit SFR); however, models with very different 
implementations of AGN feedback are showing the same problem. Alternatively, 
an excessive formation of stars in low-mass galaxies (Fontanot et al. 2009; Lo Faro 
et al. 2009; \citealt{wein11}) could lock too much gas in stars instead of ejecting 
it from DM halos and making it available for later star formation.

What is interesting to note is that, despite the very different prescriptions for
SFR and stellar feedback, all models provide similar predictions (at least up to $z$$\sim$3)
in terms of number density of objects with moderate SFRs. Probably this reflects the fact that all the 
models are calibrated to reproduce the knee of the $z$$=$0 galaxy luminosity/mass function. 
 
At low SFRs ($\la1$ M$_\odot$ yr$^{-1}$), typically below the completeness limit
of {\em Herschel} data, models 
start to separate, with R-SAM and Henriques et 
al. (2015) giving respectively the highest and lowest SFRFs.  
These differences may be due to a number of features: different modelling of 
stellar and AGN feedback, calibration procedure, treatment of merger trees 
(analytical or based on simulations).  
The turnover at low SFRs observed in all models but R-SAM is mostly due to
incompleteness: models are complete in 
DM halo mass, that is tightly correlated with 
stellar mass, while the correlation with SFR is much
broader and time-dependent, and this creates a very broad cutoff.
While at $z=0$ model SFRFs tend to be lower than the observed ones (as
noticed, e.g., by Fontanot et al. 2009), at higher redshifts (but still below $z$$\sim$1.7)
they overshoot by a large factor the SFRF of  Fontanot et al. (2012), derived for
24-$\mu$m selected sources with $M_\star>10^{10}$ M$_\odot$ in the GOODS-S
(but are consistent, up to $z$$\sim$2, with the analytic fit to our SFRF). 
Part of the difference with \citet{font12} is due to the selection in stellar mass done in that paper, 
which produce a flattening (double-peaked) of the faint-end, although the same 
paper shows that MORGANA cannot predict the drop in the SFRF at 
$\sim1$ M$_\odot$ yr$^{-1}$. 

The overprediction of models with respect to the analytic fit to data observed at $log_{10}(SFR)\lsimeq1-2$,
becoming more and more important with increasing redshift (although the faint-end slope of the observed SFRSs
is fixed at the value found at $z$$\sim$0), if confirmed by deeper data, could be due to the well known excess 
of low/intermediate mass galaxies predicted by most SAMs with respect to observed mass functions
(see, i.e., Somerville \& Primack 1999; Cole et al. 2000; Menci et al. 2002; Croton et al. 2006). 
This aspect is a result of 
the small-scale power excess typical of the $\Lambda-CDM$  power spectrum 
(e.g., \citealt{moore99}, \citealt{klyp99}, \citealt{menci12}, \citealt{cal14}). 
Some authors have suggested that a strong feedback from exploding supernovae can help 
limiting this excess, and alleviate the discrepancy between 
data and models at the smallest scales (e.g., \citealt{crot06}; \citealt{ponz12}).
 
The integrals of the SFRFs (SFRD, shown in Figure 2, {\em middle} and {\em bottom} panels) confirm and better quantify the
trends shown in Figure 1. The SFRDs obtained by integrating the SAMs over the same luminosity
range as our best-fit SFRFs, are marginally low at $z=0$ for three models out of four (the MORGANA 
prediction is very close to the observed one, but this
is driven by the bright excess visible in Fig.~1). MORGANA and the two Munich models follow the evolution 
of the observed SFRD up to $z$$\simeq$2, while the R-SAM prediction is significantly steeper,
overpredicting the data estimate from $z$$\sim$0.6 up to the higher redshifts. From
$z$$=$2 to $z$$=$3 all models but \citet{hen15} overpredict the observed SFRD (at
1-$\sigma$ uncertainty level) by a factor $>$2, with R-SAM
diverging by a factor as high as 7--8. Note that if we consider a more conservative uncertainty level
for our data (i.e., 3-$\sigma$) the disagreement would be much less or even negligible.

Since the observed SFRD shown in the {\em upper} and {\em middle} panels of Fig.~\ref{figsfrz} is obtained by integrating 
the SFRFs down to SFR values not covered by data, 
we have also computed the SFRD (from both data and models) by performing the integration just in the SFR range where 
PEP data are available (as shown in the {\em bottom} panel). Here the trend of models underpredicting the observed SFRD
at high $z$ is even more evident, with data values starting to be underestimated by SAMs at $z$$\sim$1.5--1.8, but getting closer to
model predictions (especially with the \citealt{hen15} and R-SAM ones) at $z$$\sim$3.5 (although the better agreement of the 3.2$<$$z$$<$4 SFRD might be due to the compensation of the under-estimate of the bright-end and the over-estimate of the faint-end of the SFRF).

We can conclude that our results at high-$z$ confirm a tension between models and data that possibly
points to a problem related to the common assumptions that are at the
base of galaxy formation within the $\Lambda$CDM cosmogony.
Given the good agreement between models and data at low-$z$, 
we can interpret these $z$$>$2 tensions as a consequence of model assumptions (like, e.g., the SF law and efficiency, IMF shape) and parameters being calibrated with local observations, and then assumed invariant at higher redshift. It is indeed possible for some of these analytical approximations to have an intrinsic or acquired redshift dependence (through the evolution of the physical properties of model galaxies, see, e.g., Fontanot 2014, where the predicted SFRFs for SAMs with variable IMFs are discussed). Nonetheless, the fact that the largest discrepancies between data and models are seen for the 2$<$$z$$<$3 interval, which represents a peak for both the cosmic SFR and the BH accretion, points to the treatment of SFR in extreme environments as a likely source of the tension. Indeed, the modelling of extreme environments, such those associated with the strongest starburst is still highly uncertain, with many theoretical studies suggesting a different star formation regime for these systems (see, e.g., \citealt{som01}, \citealt{hop10}).\\
On the other hand, also possible source of uncertainty in the data (affecting mainly the highest redshift interval) might contribute to the tension, increasing the bright-end of the observed SFRF:
they can be due to wrong photometric redshifts, bright IR sources at confusion level with flux enhanced by ``blending'', mis-identification of lensed galaxies (the latter only to a very minor extent, since 
the 160-$\mu$m selection should be much less affected by lensed galaxies than sub-mm wavelength ones; e.g. \citealt{neg07}). 

In the future, deeper far-IR observations will be fundamental to explore the faintest end of the SFRF, whereas a study of the bright end 
at larger redshifts will provide tighter constraints on the feedback processes regulating star formation in the brightest galaxies. 
A combined study of SFRF and mass function will be also crucial in order to fully understand and tune the single processes considered in SAMs.

\section{Conclusions}
Starting from the far-IR PEP data considered by Gruppioni et al. (2013), in this paper we investigate the evolution of the SFRF in the redshift range 0.1$<$$z$$<$4
and compare it with theoretical results from various semi-analytical models of galaxy formation (MORGANA, R-SAM, \citealt{delucia07} and \citealt{hen15}).
To compute the SFRF, we have subtracted the AGN contribution
estimated by \citet{delv14} from each SED, to obtain the IR luminosity due to SF only. Then, we have obtained the total instantaneous SFR by 
combining the IR SF luminosity with the UV derived one, to obtain the total (IR$+$UV) SFR, which we
have considered to derive the SFRF. \\
The conclusions of our work can be summarised as follows. 
\begin{itemize}
\item We find a generally good agreement between the observed and predicted SFRFs up to $z$$\sim$2 (once the observational errors are taken into account in SAMs),
with the exception of MORGANA, showing a high-SFR excess at $z$$\lsimeq$0.8.
This result implies that theoretical models, despite the different prescriptions, are able to reproduce the space-density evolution of the IR luminous galaxies from the
SFRD peak epoch up to now.
\item At $z$$\gsimeq$2, all the models start to under predict the bright-end of the SFRF. This can be due to improper modelling of several ingredients that determine 
the inflow/outflow patterns of gas in/from DM halos, like AGN feedback, limiting SFR in the largest galaxies, or even inefficient feedback from a previous generation of galaxies.
\item Our data are able to constrain the low-SFR end of the (IR$+$UV) SFRF only at low-$z$ ($z$$<$0.45). In this range of redshift, the observed slope is consistent 
with model predictions, but is steeper than the slope of the total IR LF (containing also AGN contribution) by \citet{grup13}. However, at intermediate/high-$z$, {\em Herschel} 
data do not sample the low-SFR end of the SFRF, where SAM predictions differ most: our data are not deep enough to allow us to distinguish between the different 
approaches to SF and stellar feedback considered by the different models. Additional sources of uncertainty affecting the models could be due to the fact that no evolution
with redshift is considered for local relations and functions, as SFR laws and IMF. Finally, also data at high redshift could be affected by wrong photometric redshifts and
source confusion, contributing to enhance the bright-end of the SFRF, therefore the discrepancy with model predictions.
\end{itemize}
In this work we have shown that SFRF may help putting stringent constraints on the physical processes modelled in SAMs, especially if extended to low-SFRs, while
a study of the bright-end at larger redshifts will provide tighter constraints on the feedback processes regulating star formation in the brightest galaxies. 
A combined study of SFRF and mass function will be crucial in order to fully understand and tune the single processes modelled in SAMs and to have a global picture 
of the evolution of SFR and mass growth in galaxies.

\section*{Acknowledgments}
CG and FP acknowledge financial contribution from the contracts 
PRIN-INAF 1.06.09.05 and ASI-INAF I\/005\/07/1 and I\/005\/11\/0.
FF acknowledges financial support from the grants PRIN MIUR 2009 ``The Intergalactic
Medium as a probe of the growth of cosmic structures'' and PRIN INAF 2010 ``From the
dawn of galaxy formation''.

\bsp

\label{lastpage}

\end{document}